\documentclass[english,aps,prx,twocolumn,superscriptaddress,longbibliography]{revtex4-2}
\usepackage{amssymb}
\usepackage{amsmath}
\usepackage{graphicx}
\usepackage{natbib}
\usepackage{epstopdf}
\usepackage{color}
\usepackage{braket}
\usepackage{physics}
\usepackage{mathrsfs}
\usepackage{upgreek}
\usepackage{todonotes}
\usepackage[colorlinks=true, allcolors=blue]{hyperref}
\usepackage{soul}
\usepackage{ulem}

\definecolor{mg}{rgb}{0,0,.5}

\newcommand{\UCLA}{\affiliation{Department of Physics and Astronomy, University of California -- Los Angeles, Los Angeles, California 90095, USA}}
\newcommand{\CQSE}{\affiliation{Center for Quantum Science and Engineering, University of California -- Los Angeles, Los Angeles, California 90095, USA}}
\newcommand{\ECE}{\affiliation{Department of Electrical and Computer Engineering, University of California, Los Angeles, California 90095, USA}}
\newcommand{\princeton}{\affiliation{Department of Physics, Princeton University, Princeton, New Jersey~08544, USA}}
\newcommand{\UMD}
{\affiliation{Joint Center for Quantum Information and Computer Science, NIST/University of Maryland, College Park, Maryland 20742, USA}}
\newcommand{\NIST}{\affiliation{National Institute of Standards and Technology, Gaithersburg, MD 20899, USA}}
\begin{document}
\title{Effects of residual exchange coupling on simultaneously driven spin qubits}

\author{Heun Mo Yoo}
\UCLA

\author{Tanner M. Janda}
\UCLA

\author{Victor Yu}
\UCLA

\author{Michael J. Gullans}
\UMD
\NIST

\author{Adam R. Mills}
\princeton

\author{Jason R. Petta}
\email{petta@physics.ucla.edu}
\UCLA
\CQSE
\ECE

\begin{abstract}
Exchange coupling and microwave drives are widely used control mechanisms for spin qubits. However, the influence of exchange coupling on microwave-driven spin dynamics is not fully understood. We report simultaneous drive measurements of two spin qubits as a function of exchange coupling and drive power. When the Rabi frequency exceeds exchange, we observe a beating pattern in the Rabi oscillations. In the opposite limit, two exchange-split patterns emerge in the Rabi chevrons. We find that these exchange-induced effects are suppressed when the difference in the Rabi frequencies exceeds the exchange coupling. A theoretical model is developed that reproduces the main features in our data.
\end{abstract}

\maketitle
The exchange interaction is the primary entangling mechanism for electron spin qubits \cite{Loss1998,burkardSemiconductorSpinQubits2023}. Since the first demonstration of coherent control of electron spins using exchange in a semiconductor quantum device \cite{Petta2005}, exchange coupling has been applied to a variety of spin qubits, such as Loss-DiVincenzo (LD) single spin qubits \cite{Loss1998}, two-electron singlet–triplet spin qubits \cite{Shulman2012,Nichol2017}, and three-electron exchange-only spin qubits \cite{Medford2013,Medford2013b,Eng2015}. In LD spin qubits, for example, exchange coupling between neighboring spins is used to perform two-qubit gates \cite{Veldhorst2015}, while single-spin manipulation is achieved with electric dipole spin resonance (EDSR) \cite{Nowack2007,Pioro2008}.

A growing number of teams have demonstrated high-fidelity gate operations with exchange coupling and EDSR \cite{mills_2Q_2021,xue2022quantum,Noiri2022}. Despite these achievements, the influence of exchange coupling on simultaneously driven spins has not been thoroughly investigated. It remains unclear how residual exchange coupling, present even in the absence of exchange pulses, affects parallelized gate operations. For example, in a recent experiment reporting record-high single qubit gate fidelities ($F$ $>$ 99.999\%), the exchange coupling was kept below 10 kHz to minimize its impact on single qubit gates \cite{takeda2026}. However, the vanishingly small exchange coupling used in that experiment may not be applicable to implementations requiring fast two-qubit gates, particularly when the on-off ratio of the exchange interaction is constrained \cite{Philips2022,Takeda2022}. Moreover, in prior experiments \cite{xue_benchmarking_2019,Undseth2023}, heating and off-resonant excitations from microwave pulses were the dominant sources of qubit frequency shifts, making it difficult to isolate the effects of exchange coupling on driven spin dynamics.

In this Letter, we investigate the time-evolution of two simultaneously driven spin qubits as a function of the qubit Rabi frequencies and exchange coupling. Our results show that when exchange is much less than the Rabi frequencies, a beating pattern emerges in the Rabi oscillations, with a beat frequency proportional to exchange. On the other hand, when exchange is greater than the Rabi frequencies, we observe two partially overlapping Rabi chevrons that are separated by exchange. Moreover, the drive power dependence indicates that the spins can undergo independent rotations when the Rabi frequency difference exceeds exchange. Based on these findings, we construct a phase diagram illustrating three different regimes of entangling dynamics and explain how these dynamics impact simultaneous single qubit gate operations.

\begin{figure}[t!]
	\centering
    \includegraphics[width=\columnwidth]{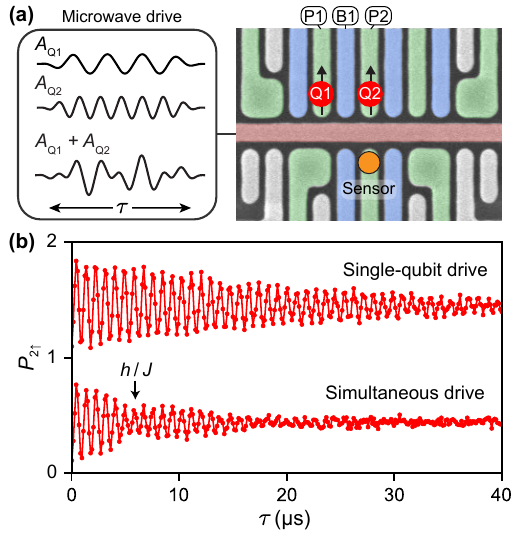}
	\caption{Device and Rabi oscillations. (a) False-colored scanning electron microscope image of a similar device. Two spins, Q1 and Q2, are confined under plunger gates P1 and P2. The exchange coupling is tuned by the interdot barrier gate B1. A raised-cosine shaped microwave pulse with duration $\tau$ is applied to the middle screening gate \cite{xue2022quantum,Takeda2022}. $A_{\text{Q1}}$ ($A_{\text{Q2}}$) denotes the drive amplitude for Q1 (Q2). A charge sensor located below the central screening gate is used for qubit readout \cite{mills2022high}. (b) Rabi oscillations measured when driving Q2 only (top) and simultaneously driving both qubits (bottom). $P_{2\uparrow}$ represents the spin-up probability of Q2. When both qubits are driven simultaneously, $P_{2\uparrow}$ exhibits a beating pattern, with the first node at ${\tau} = h/J$ (see the black arrow). The traces are vertically offset by one for clarity.
    }
	\label{fig:1}
\end{figure}

We carry out our experiment using an Intel triple quantum dot (TQD) \cite{George2025}. Figure~\ref{fig:1}(a) shows the gate electrode pattern. Two qubits, labeled Q1 and Q2, are defined in the Si quantum well beneath plunger gates P1 and P2. To initialize and read out the spin states of Q1 and Q2, we use a combination of energy dependent tunneling with the left Fermi reservoir and spin shuttling~\cite{Janda2026}. An in-plane magnetic field $B$~=~337~mT is applied, resulting in qubit resonance frequencies $f_{\rm Q1}$~=~11.564~GHz and $f_{\rm Q2}$~=~11.622~GHz. EDSR is achieved by applying microwaves to the central screening gate, resulting in Rabi frequencies $\Omega_{\text{Q1}}$ and $\Omega_{\text{Q2}}$ \cite{George2025}.

We first characterize single qubit Rabi oscillations by driving Q1 and Q2 individually. The upper curve in Fig.~\ref{fig:1}(b) shows the spin-up probability of Q2, $P_{2 \uparrow}$, as a function of the microwave pulse length, $\tau$, with only Q2 being driven. By fitting the Rabi oscillations to a damped sinusoid we extract $\Omega_{Q2}$~=~1.1~MHz \cite{Takeda2016}. 

To investigate the dynamics of two simultaneously driven spin qubits, we apply drive tones at $f_{\rm Q1}$ and $f_{\rm Q2}$ with the drive amplitudes tuned such that $\Omega_{Q1}$ $\approx$ $\Omega_{Q2}$~=~1.1~MHz.  We observe pronounced changes in the Rabi oscillations, as illustrated by the lower trace in Fig.~\ref{fig:1}(b).  First, we observe a beating pattern in $P_{2 \uparrow}$($\tau$), with the first node located at $\tau\approx$~6~{\textmu}s. The node occurs on a timescale $\sim h/J$, where $h$ is Planck's constant and $J/h\approx160~\text{kHz}$ is the exchange coupling frequency deduced from decoupled CZ oscillations \cite{Watson2018,xue2022quantum}. Second, we observe a more rapid decay of the Rabi oscillations due to the increased power of the combined microwave pulses heating the device. Since the Rabi frequency remains unchanged, we can exclude a scenario where heating shifts the qubit resonance frequencies and alters the Rabi oscillations during simultaneous drive \cite{Undseth2023,xue_benchmarking_2019}.

\begin{figure}[t!]
	\centering
    \includegraphics[width=\columnwidth]{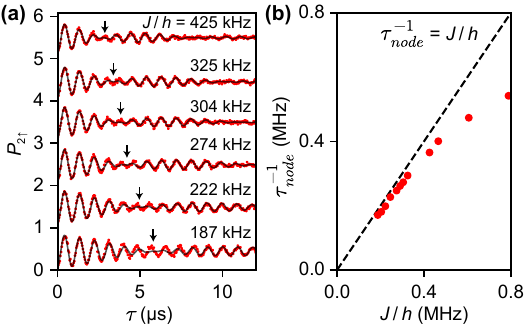}
	\caption{Exchange coupling dependence of two-spin driven dynamics. (a) Rabi oscillations of Q2 measured at different values of $J$ while both qubits are simultaneously driven with $\Omega_{\text{Q1}} = \Omega_{\text{Q2}}$~=~1.1~MHz. The data (red dots) are fit (black lines) with a damped sinusoid that is modulated by an additional low-frequency component. With increasing $J$, $\tau_{node}$, defined as the first node location in $\tau$, shifts to shorter times (see the black arrow). The traces are vertically offset by one for clarity. (b) $\tau_{node}^{-1}$ as a function of $J/h$. The dashed line represents the trend  $\tau_{node}^{-1}=J/h$ inferred from the data at weak coupling ($J/h\ll\Omega_\text{Q1},\Omega_\text{Q2}$).}
	\label{fig:2}
\end{figure}

\begin{figure*}[tb]
	\centering
    \includegraphics[width=2\columnwidth]{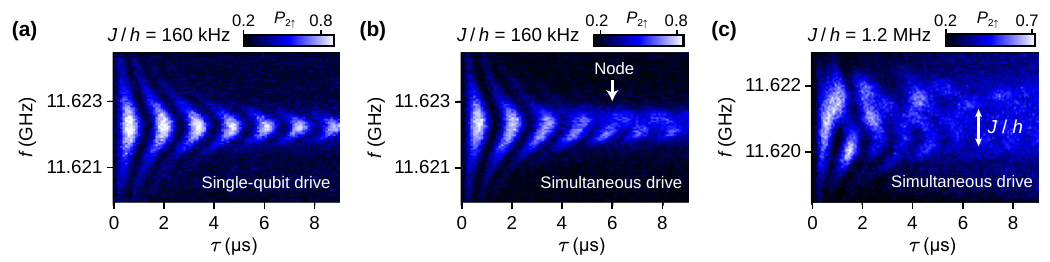}
	\caption{Crossover from weak to strong exchange coupling. (a) Q2 Rabi chevron measured when driving Q2 only. (b) Q2 Rabi chevron measured when simultaneously driving Q1 and Q2. A node appears at $\tau=h/J$. (c) Q2 Rabi chevron measured when simultaneously driving Q1 and Q2 with strong $J/h$~=~1.2~MHz. Intricate patterns emerge at two different drive frequencies (see white arrow) due to the exchange splitting \cite{Veldhorst2015,Zajac2017CNot}. The Q1 drive frequency is fixed at the $\ket{\downarrow\downarrow} \leftrightarrow \ket{\uparrow\downarrow}$ resonance while the Q2 drive frequency is swept across the $\ket{\downarrow\downarrow} \leftrightarrow \ket{\downarrow\uparrow}$ and $\ket{\uparrow\downarrow} \leftrightarrow \ket{\uparrow\uparrow}$ resonances.}
   	\label{fig:3}
\end{figure*}

We more broadly explore the effect of exchange coupling on simultaneously driven Rabi oscillations by dynamically tuning $J$ with a barrier gate pulse [gate B1 in Fig.~\ref{fig:1}(a)]. During the barrier pulse, we compensate the plunger gate voltages $V_{\rm P1}$ and $V_{\rm P2}$ to keep the system at the symmetric operating point \cite{Reed2016,Martins2016}. For each exchange pulse amplitude, we adjust the microwave drive frequencies to drive both qubits on resonance. Figure~\ref{fig:2}(a) shows $P_{2 \uparrow}(\tau)$ measured at different values of $J$.  For the values of $J$ investigated, we again observe beating patterns in the Rabi oscillations. Furthermore, as $J$ increases, the first node shifts to smaller $\tau$, indicating an approximately inverse relationship between the node position and $J$.

We analyze the data by fitting $P_{2 \uparrow}(\tau)$ to a damped sinusoidal oscillation modulated by a low-frequency sinusoidal envelope [see the black lines in Fig.~\ref{fig:2}(a)]:
\begin{equation}
    P_{2 \uparrow}(\tau) = A \cos\left(2\pi\Omega \tau + \phi\right) \cos\left(\frac{\pi}{2} \frac{\tau}{\tau_{node}}\right) e^{-
    \tau/T} + C,
\label{eq1}
\end{equation}
where $A, \Omega, \phi, \tau_{node}, T,$ and $C$ are fitting parameters. In the fit, the first node of the envelope occurs at $\tau=\tau_{node}$. Figure~\ref{fig:2}(b) shows the extracted $\tau_{node}^{-1}$ as a function of the $J/h$ values obtained by measuring decoupled CZ oscillations \cite{Watson2018,xue2022quantum}. We find that when $J/h$~$\ll$~$\Omega_{Q1}$~$\approx$~$\Omega_{Q2}$, $\tau_{node}^{-1}$ is closely approximated by $J/h$. The relation $\tau_{node}^{-1}\approx J/h$ corroborates our earlier observation that residual coupling affects simultaneously driven spins on the timescale of $h/J$. On the other hand, when $J/h$~$\gtrsim$~$\Omega_{Q1}$~$\approx$~$\Omega_{Q2}$, $\tau_{node}^{-1}$ deviates from the linear dependence, suggesting that more complex dynamics emerge at increased coupling.

To explore the crossover to the strong coupling regime ($J/h \gg \Omega_{\text{Q1}},\Omega_{\text{Q2}}$), we reduce the Rabi frequencies to $\Omega_{Q1}$~$\approx$~$\Omega_{Q2}$~=~0.74~MHz. As a starting point, we characterize the drive frequency dependence of Q2 when only it is driven. Consistent with expectations, we observe the expected Rabi chevron centered at the resonance frequency of Q2 [see Fig.~\ref{fig:3}(a)] \cite{Kawakami2014,Veldhorst2014}. Next, we simultaneously drive both qubits without applying an exchange pulse. In the absence of exchange pulses, the two qubits are in the weak coupling regime, with an exchange coupling $J/h$~=~160~kHz that is much smaller than the Rabi frequencies. The overall chevron pattern is similar to Fig.~\ref{fig:3}(a), except for the damping of the oscillations at $\tau\approx6$~{\textmu}s [see Fig.~\ref{fig:3}(b)]. The value of $\tau$ at which the damping occurs is close to $h/J$, in agreement with the node observed in Fig.~\ref{fig:2}. Lastly, we repeat the measurement in the strong coupling regime. We set $J/h$~$\gg$~$\Omega_{Q1}$~$\approx$~$\Omega_{Q2}$ by applying an exchange pulse while simultaneously driving Q1 and Q2. The Rabi chevron measured at strong coupling is markedly different from the typical single-qubit chevron pattern [see Fig.~\ref{fig:3}(c)]. For short drive durations $\tau<$~2~{\textmu}s, we identify two chevron-like patterns centered at $f\sim$~11.620~GHz and 11.621~GHz. The separation in frequency of these two patterns corresponds to $J/h$. 

In the strong coupling regime, the exchange coupling gives rise to two distinct resonance frequencies in the Rabi chevron \cite{Zajac2017CNot}. One resonance occurs when the two qubits are driven at the $\ket{\downarrow\downarrow} \leftrightarrow \ket{\uparrow\downarrow}$ and $\ket{\downarrow\downarrow} \leftrightarrow \ket{\downarrow\uparrow}$ transitions, with each drive frequency $J/2h$ below the uncoupled resonance frequency. Since the two drive tones are in phase, they induce a transition between the $\ket{\downarrow\downarrow}$ and $\ket{\uparrow\downarrow}+\ket{\downarrow\uparrow}/\sqrt{2}$ states. Another resonance appears when the frequency of the drive applied to one of the qubits, for example Q2, is increased by $J/h$. In this case, Q1 and Q2 are driven at the $\ket{\downarrow\downarrow} \leftrightarrow \ket{\uparrow\downarrow}$ and $\ket{\uparrow\downarrow} \leftrightarrow \ket{\uparrow\uparrow}$ transitions, respectively, leading to oscillatory dynamics involving the $\ket{\downarrow\downarrow}$, $\ket{\uparrow\downarrow}$, and $\ket{\uparrow\uparrow}$ states \cite{Supplement}.

Having characterized the Rabi chevrons in the weak and strong coupling regimes, we explore how a difference in Rabi frequencies ($\Omega_{\text{Q1}} \neq \Omega_{\text{Q2}}$) affects the Rabi oscillations in the weak coupling regime. We set the  microwave drive tones at each of the qubits' uncoupled resonance frequencies. Figure~\ref{fig:4}(a) shows the Q2 Rabi oscillations measured as a function of the Rabi frequency difference, $\Omega_{-}=\Omega_{\text{Q2}}-\Omega_{\text{Q1}}$, at $J/h$~=~160~kHz. When $|\Omega_{-}|$ is large, corresponding to small and large values of 
$\Omega_{\text{Q1}}$, we observe Rabi oscillations similar to those obtained during single qubit driving. The node and beating pattern are absent even in the presence of residual exchange coupling. On the other hand, with the Q1 drive amplitude tuned so that $\Omega_{\text{Q1}}$~$\approx$~$\Omega_{\text{Q2}}$, we observe a node in the oscillations at $\tau=$~6~{\textmu}s, consistent with the relation $\tau=h/J$ expected in the weak coupling regime. This observation suggests that the effect of exchange coupling appears only when the two Rabi frequencies are comparable.

In the presence of a large magnetic field gradient, a system of two spins is well described by the parallel and anti-parallel spin basis \{$\ket{\downarrow\downarrow}$, $\ket{\uparrow\downarrow}$, $\ket{\downarrow\uparrow}$, $\ket{\uparrow\uparrow}$\}, with the energies of the anti-parallel states lowered by the exchange interaction \cite{Zajac2017CNot}. To more clearly highlight the effects of exchange coupling, we consider two qubits driven at their uncoupled resonance frequencies and express the Hamiltonian in the basis of the Bell states {$\ket{\Phi^{\pm}}=(\ket{\uparrow\uparrow}\pm\ket{\downarrow\downarrow})/\sqrt{2}$ and {$\ket{\Psi^{\pm}}=(\ket{\uparrow\downarrow}\pm\ket{\downarrow\uparrow})/\sqrt{2}$. In the rotating frame, the Hamiltonian in the \{$\ket{\Phi^{+}} $,$\ket{\Psi^{+}} $,$\ket{\Phi^{-}} $,$\ket{\Psi^{-}} $\} basis takes the form
\begin{equation}
H = \frac{h}{2}\begin{bmatrix}
0                   & \Omega_{+} & 0 &    0\\ 
\Omega_{+}  & -J/h               & 0                  &0\\ 
0  & 0                  & 0               & \Omega_{-}\\ 
0                   & 0 & \Omega_{-} & -J/h
\end{bmatrix},
\end{equation}
where $\Omega_{+}=\Omega_{\text{Q1}}+\Omega_{\text{Q2}}$ denotes the sum of the Rabi frequencies. In our experiment, the qubits are driven around the same axis of rotation with positive $\Omega_{\text{Q1}}$ and $\Omega_{\text{Q2}}$, constraining the parameter space to $|\Omega_{-}|\leq\Omega_{+}$. Therefore, we define dimensionless parameters $h\Omega_{+}/J$ and $|h\Omega_{-}/J|$ that determine three distinct regimes of simultaneous driving: (1) weak coupling regime, (2) strong coupling regime, and (3) nearly independent regime.

\begin{figure}[htb!]
	\centering
    \includegraphics[width=\columnwidth]{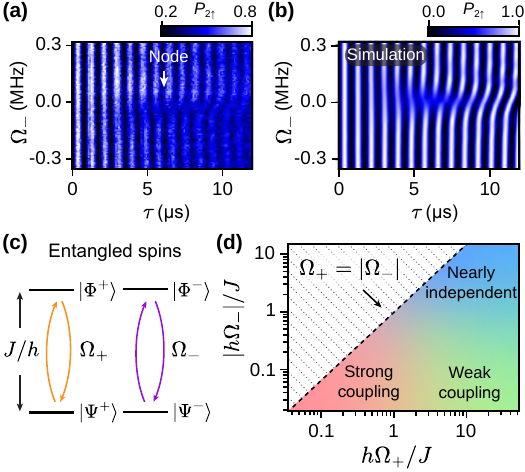}
	\caption{Effect of Rabi frequency difference and entangled-spin model. (a) Rabi oscillations of Q2 measured as a function of $\Omega_{-}$. $\Omega_{\text{Q1}}$ varies linearly along the vertical axis, whereas $\Omega_{\text{Q2}}$ is fixed at 1.1~MHz. The node indicated by the white arrow disappears upon tuning the Rabi frequency difference between Q1 and Q2. (b) Simulated $\Omega_{-}$ dependence of Rabi oscillations. The simulation is performed with $J/h$~=~160~kHz and $\Omega_{\text{Q2}}$~=~1.1~MHz while varying $\Omega_{\text{Q1}}$. (c) Energy level diagram in the \{$\ket{\Phi^{+}} $,$\ket{\Psi^{+}} $,$\ket{\Phi^{-}} $,$\ket{\Psi^{-}} $\} basis. $\Omega_{+(-)}$ represents a transition rate between the $\ket{\Psi^{+(-)}}$ and $\ket{\Phi^{+(-)}}$ states, split by $J/2$.  (d) Phase diagram illustrating three regimes of simultaneous driving. The hatched region is forbidden for positive $\Omega_{\text{Q1}}$ and $\Omega_{\text{Q2}}$. When both are positive, the strength of $\Omega_{+}$ compared to $J$ determines the crossover between the strong and weak coupling regimes. In the weak coupling regime ($|\Omega_+| \gg J/h$), a second crossover occurs at $h|\Omega_-|/J \gg 1$, where the two spins undergo nearly independent Rabi oscillations.}
	\label{fig:4}
    
\end{figure}

(1) At large $h\Omega_{+}/J$ but small $|h\Omega_{-}/J|$, the system is in the weak coupling regime, where a beating pattern appears in the Rabi oscillations. Starting from the initial state $\ket{\downarrow\downarrow} = \frac{1}{\sqrt{2}} \left(\ket{\Phi^+} - \ket{\Phi^-}\right)$, simultaneously applied microwave fields drive transitions between $\ket{\Phi^+}$ and $\ket{\Psi^+}$. During the driven evolution, $J$ induces slow superimposed dynamics as the $\ket{\Phi^+}$ and $\ket{\Psi^+}$ states accumulate a phase relative to the weakly driven $\ket{\Phi^-}$ component. For the simple case of $\Omega_- = 0$, a maximally entangled state forms, which manifests as a node in the Rabi oscillations at $\tau=h/J$.

(2) At small $h\Omega_{+}/J$ and $|h\Omega_{-}/J|$, the system enters the strong coupling regime. In this case, $J/h$ exceeds the EDSR linewidth, which scales with the Rabi frequency. Thus, we can resolve two overlapping chevron patterns split by $J$. To validate our model, we simulate the Rabi chevrons in the strong coupling regime and confirm good agreement with the experimental data \cite{Supplement}.

(3) At large $|h\Omega_{-}/J|$, the time evolution of the spins is nearly independent. When the spins are driven with different Rabi frequencies such that $|\Omega_{-}|$~$>$~$J/h$, both the $\ket{\Phi^{+}} \leftrightarrow \ket{\Psi^{+}}$ and $\ket{\Phi^{-}} \leftrightarrow \ket{\Psi^{-}}$ transitions occur. Driving these transitions together suppresses the node at $\tau=h/J$ because the interference between the positive-phase states ($\ket{\Phi^{+}}$ and $\ket{\Psi^{+}}$) and their negative-phase counterparts ($\ket{\Phi^{-}}$ and $\ket{\Psi^{-}}$) prevents the formation of the maximally entangled state. Based on this model, a crossover from entangled spins to nearly independent spins occurs as $|\Omega_{-}|$ increases and becomes larger than $J$ [see the simulation in Fig.~\ref{fig:4}(b)].

In summary, we have characterized the impact of the exchange interaction on two simultaneously driven spin qubits. By tuning the exchange coupling, as well as the amplitudes and frequencies of microwave drives, we identify three regimes of simultaneous driving in which exchange coupling affects qubit rotations to varying degrees. At strong coupling, the exchange interaction splits the qubit resonance frequencies, giving rise to intricate Rabi chevron patterns. At weak coupling, the qubit resonance frequencies remain unchanged, but slow entangling dynamics produce beating patterns in the Rabi oscillations. With a large Rabi frequency difference, the spins undergo nearly independent Rabi oscillations.

These observations suggest several ways to mitigate the effects of residual exchange coupling in quantum dot arrays. First, residual exchange coupling must be reduced below $\Omega_{+}$ to place the system in the weak coupling regime. As long as gate operations are carried out much faster than $h/J$, the influence of entangling dynamics on gate performance is negligible. Second, if the speed of gate operations or the dynamic control over exchange coupling is limited, then the qubits should be driven at different Rabi frequencies. With $|\Omega_{-}|>J/h$, unwanted exchange-induced dynamics that degrade parallel gate operations can be suppressed even when the qubits have appreciable coupling strength. While we expect that these approaches will improve the parallel control of spin qubits, further studies on simultaneous drives with varying phases and two-qubit gates will provide additional insight into the effects of residual exchange coupling in dense spin qubit arrays \cite{HRL2026}.

\begin{acknowledgments}
Research was sponsored by Army Research Office Grant No.~W911NF-23-1-0104.  Intel devices were provided through the Qubits for Computing Foundry under Cooperative Agreement No.~W911NF-22-2-0037.  The views and conclusions contained in this document are those of the authors and should not be interpreted as representing the official policies, either expressed or implied, of the Army Research Office or the U.S. Government. The U.S. Government is authorized to reproduce and distribute reprints for Government purposes notwithstanding any copyright notation herein.
\end{acknowledgments}

\bibliography{RMP_master2_bib_v8}

\end{document}